# Comprehensive Anisotropic Linear Optical Properties of Weyl Semimetals, TaAs and NbAs

Rui Zu,[1] Mingqiang Gu,[6,*] Lujin Min,[1] Chaowei Hu,[5] Ni Ni,[5] Zhiqiang Mao,[2] James M. Rondinelli,[4] Venkatraman Gopalan[1,2,3,†]

[1]Department of Materials Science and Engineering, Pennsylvania State University, University Park, Pennsylvania, 16802, USA

[2]Department of Physics, Pennsylvania State University, University Park, Pennsylvania, 16802, USA

[3]Department of Engineering Science and Mechanics, Pennsylvania State University, University Park, Pennsylvania, 16802, USA

[4]Department of Materials Science and Engineering, Northwestern University, Evanston, Illinois, 60208, USA

[5]Physics and Astronomy Department, University of California, Los Angeles, Los Angeles, California, 90095, USA

[6]Shenzhen Institute for Quantum Science and Engineering (SIQSE) and Department of Physics, Southern University of Science and Technology, Shenzhen 518055, China

Abstract:

TaAs and NbAs are two of the earliest identified Weyl semimetals that possess many intriguing optical properties, such as chirality-dependent optical excitations and giant second harmonic generation (SHG). Linear and nonlinear optics have been employed as tools to probe the Weyl physics in these crystals. Here we extend these studies to address two important points: determining the complete anisotropic dielectric response, and to explore if and how they can reveal essential Weyl physics. For the first time, we determine the complete anisotropic dielectric functions of TaAs and NbAs by combining spectroscopic ellipsometry and density functional theory (DFT). Parameterized Lorentz oscillators are reported from 1.2-6 eV (experiment) and 0-6 eV (DFT), and good agreement is shown between them. Both linear and nonlinear optical properties have been reported to reveal Weyl physics. We suggest that strong optical resonances from trivial bands are the likely origin of the large optical second harmonic generation previously reported at these energies. Furthermore, by comparing the contribution of a small $k$-space centered around the Weyl cones to the total linear dielectric function, we find that these contributions are highly anisotropic and are <25% of the total dielectric function below 0.5 eV; above 1eV, these contributions are negligible. Thus, the study of Weyl physics using optical techniques requires very low energies and even there, a careful assessment is required in distinguishing the much smaller contributions of the Weyl bands from the dominant contributions of the trivial bands and Drude response to the total dielectric function.

*gumq@sustech.edu.cn
†vxg8@psu.edu

## I. INTRODUCTION

Although Weyl fermion was initially predicted in high energy physics [1], it was first experimentally discovered in condensed matter, namely the TaAs family [2]. Weyl fermions can be described as low-energy excitations near the Weyl nodes which represent the crossing points of singly-degenerate bands in the presence of broken inversion symmetry or broken time-reversal symmetry. Weyl nodes appear in pairs with opposite chiralities; the bulk and surface state correspondence results in surface Fermi arcs. All these hallmarks of Weyl states have been demonstrated experimentally [3–10]. Weyl fermions give rise to a wide range of exotic transport properties [11], such as extremely high mobility [12], negative longitudinal magnetoresistance caused by chiral anomaly [13], and intrinsic anomalous Hall effect [14–17].

Optical measurements have been used as a powerful technique to characterize the exotic properties of Weyl semimetal states. [18–22] To explore the linear band crossing near the Fermi level, the temperature-dependent Drude behavior in the low energy range and linear optical conductivity near the Weyl points have been explored. [22–24] The optical transitions in TaAs near the two types of Weyl nodes (labeled W1 and W2) have been identified to be <0.1 eV (for W1) and <0.2 eV (for W2). [23] Due to the intrinsic noncentrosymmetry in the TaAs family, strongly anisotropic second harmonic (SHG) and circular photo-galvanic effect (CPGE) have been observed, prompting further interest in higher-order nonlinear and coupled optical properties. [18,25] The giant SHG response in the TaAs family has been attributed to linear resonances, as well as the shift current. [25–28] While extensive studies have focused on exploring the spectroscopic behavior at energies near Weyl points [21–23], and the optical resonances that contribute to giant SHG [26], the complete *anisotropic* linear optical susceptibility tensor,

*anisotropic* complex refractive indices, and the band- and momentum- resolved optical transitions in TaAs and NbAs across the visible and near-infrared spectrum remain unreported. To that end, we report the complete anisotropic linear optical properties of both TaAs and NbAs, and discuss the consequences of anisotropy in utilizing optical probes for probing Weyl physics.

Previous studies on the natural (112) planes of the crystal were focused on optical second harmonic generation, THz emission, and pump-probe measurements [20,25,26,29]. Studies on (001) crystal surface focused on angle-resolved photoemission spectroscopy (ARPES) and linear spectroscopy at low energies from 0-1eV [2,7,23,30,31]. Linear spectroscopy and Raman studies also exist on (107) orientation [21,22]. In this work, we use the (010) crystal plane, which allows us to directly probe properties along the principal axes [100] and [001] in the crystal plane, and study the anisotropic linear optical properties and resonances in TaAs and NbAs. We also studied (112) planes to confirm the uniqueness of the measured dielectric function. Large anisotropic optical dielectric resonances are found below 3.5 eV. The complete parameterized Lorentz oscillators within 1.2-6 eV (experiment, Expt.) and 0-6 eV (DFT) are reported, corresponding to interband transitions. Strong resonances are found below 1 eV between the first two conduction and valence bands near the Fermi level, which originate from Ta ($5d$) to Ta ($5d$) and As ($4p$) to Ta ($5d$) transitions. Resonances are observed near the energies where previous optical second harmonic generation (SHG) experiments were performed [25,26]; they are expected to enhance the SHG effect. The contribution to the dielectric function from a localized $k$ space volume (~0.0125% of the volume of the 1$^{st}$ Brillouin zone) centered around the Weyl points is also quantified; we find that these contributions are <25% of the total dielectric function below 0.5 eV, and are negligible at energies greater than 1 eV, indicating low-energy optical probes are required to discern topological band features in the TaAs family.

## II. METHODS

TaAs and NbAs exhibit a similar crystal structure (**Fig. E1(a)**) and the same point group $4mm$ ($a = 3.4348$Å, $c = 11.641$Å for TaAs; $a = 3.452$Å, $c = 11.679$Å for NbAs; $c$-axis in both cases is parallel to the [001] direction). [32,33] Single crystals of TaAs and NbAs were grown by chemical vapor transport with stoichiometrically mixed Ta (or Nb) and As powders at around 1000°C for four weeks (details can be found in **Appendix E**). As-grown surfaces of (020) are determined by $\theta - 2\theta$ X-ray diffraction confirming single-crystalline properties and (020) out-of-plane direction (See **Fig. E2**). The in-plane [001] and [100] directions are then determined by Laue backscattering diffraction and confirmed by the electron backscatter diffraction (EBSD) (See **Fig. E3**).

Crystal quality plays a vital role in the intrinsic properties such as vacancies and stacking faults, which could shift the positions of Weyl nodes and alter the Fermi surface [34]. A previous study [35] shows that the Weyl nodes move closer to the Fermi level as the mobility increases. The quality of the studied crystals is examined using Hall measurements (**Fig. F1**). Analysis of the transport data show that the electron mobility of TaAs is $\mu_e(10K) = 6.04 \: m^2V^{-1}s^{-1}$ and for the NbAs to be $\mu_e(10K) = 83.5 \: m^2V^{-1}s^{-1}$, which is comparable to that reported in the literature. [36–39] Thus, the above properties suggest that the crystals studied here are reasonable representatives of the crystals reported in the literature.

Spectroscopic ellipsometry is performed using Woollam M-2000F focused beam spectroscopic ellipsometer with wavelengths from 1000 nm (1.2 eV) to 200 nm (6 eV). Amplitude change and phase shift of the specularly reflected light are collected as a function of the wavelength, which is sensitive to the crystal orientations. Three different orientations (see below) are measured and fitted simultaneously with the Lorentz model to confirm the uniqueness of the model. The complex dielectric permittivity tensor for the uniaxial system is given by the diagonal second rank tensor, $\tilde{\varepsilon} = (\tilde{\varepsilon}_{11} = \tilde{\varepsilon}_{22}, \tilde{\varepsilon}_{33})$, where subscripts 1 and 3 denote crystallographic *a* and *c* axis. The three different crystal cuts used in this study are represented as orientation **A**: [001] // lab *x*, [010] // lab *z*; orientation **B**: [001]// lab *y*, [010]// lab *z* and orientation **C**: [1 -1 0]// lab *x*, (112)// lab *z* respectively where the *x-z* plane is the plane of incidence. Kramers-Kronig relationship between real and imaginary components of the dielectric tensor is imposed during the modeling of the ellipsometry data. The imaginary component of the dielectric function is forced to be positive in all the fittings.

Density functional theory (DFT) calculations are performed using Vienna Ab-initio Simulation Package (VASP) [40] to provide insights into the electronic properties leading to the optical dielectric function. The Perdew-Burke-Ernzerhof (PBE) functionals [41] with spin-orbital coupling (SOC) interactions are used in all calculations with a planewave expansion up to 500 eV. The projector augmented wave (PAW) method [42] is used to treat the core and the valence electrons using the following electronic configurations: $5p^66s^25d^3$ for Ta, $4p^65s^24d^3$ for Nb, and $4s^24p^3$ for As. A 30×30×30 Γ-centered Monkhorst-Pack *k*-point mesh is used for calculating the electron density of states (DOS) and the optical properties. Multiple smearing factors have been tested from 0.02 eV to 0.2 eV to find the best match to the experimental results. Two representative smearing factors (0.02 and 0.2 eV) are then evaluated, and since the results from the 0.2 eV smearing factor matched the experiments better, these are presented in the main text (See **Fig. G1-3** and **Table. GI-IV** for results using a smearing factor of 0.02 eV). To evaluate the contribution of the dielectric function from the momentum space near the Weyl points, we used dense *k*-point grids and calculated the dielectric function contribution from a small volume $\frac{(\vec{b}_1 \times \vec{b}_2) \cdot \vec{b}_3}{20^3}$ centered around either WP1 or WP2. Note that $\vec{b}_1$, $\vec{b}_2$ and $\vec{b}_3$ are reciprocal lattice parameters for the primitive cell. Due to the metallic nature of TaAs and NbAs, an isotropic free carrier response [43] is also added to the calculated dielectric function, and the as sociated coefficients are included in **Tables. AI-IV.**

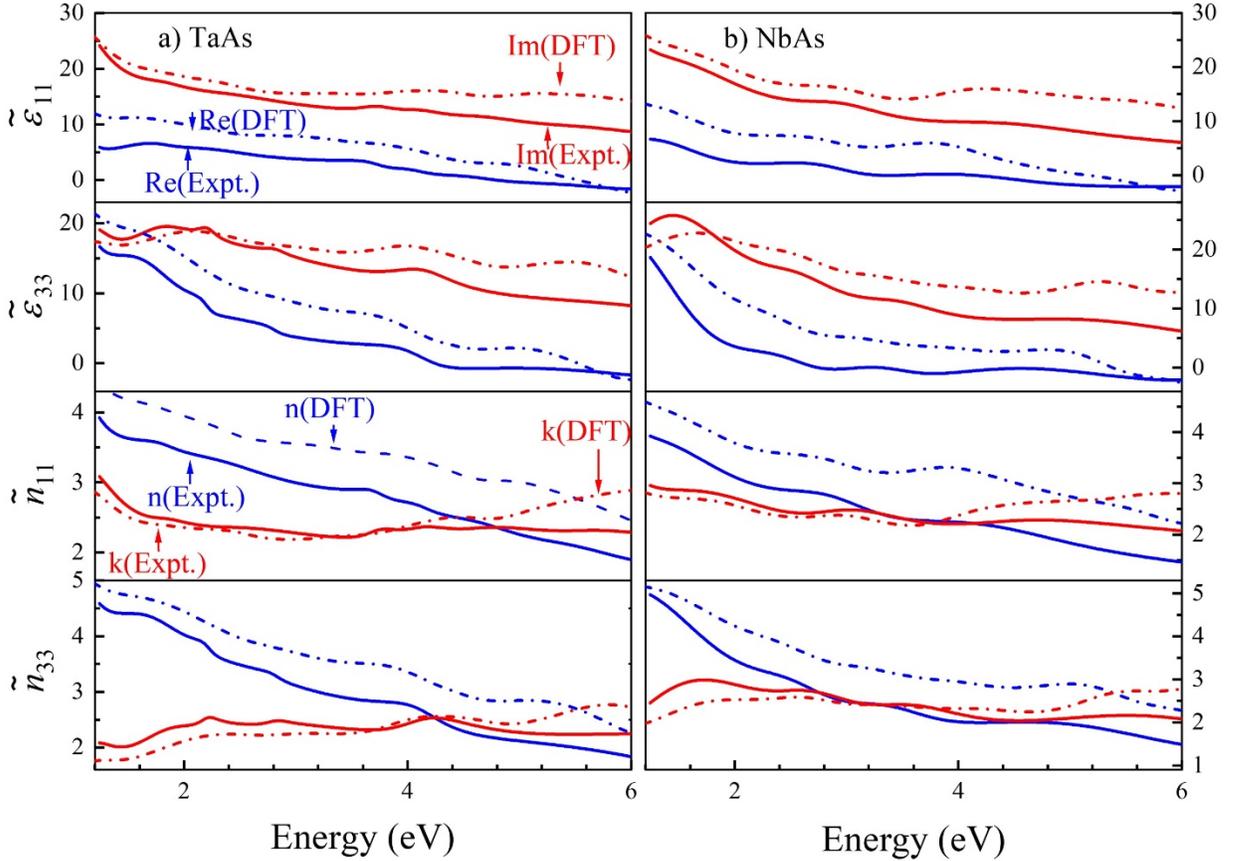

FIG 1. The complex dielectric functions $\tilde{\varepsilon}_r$ and refractive indices $\tilde{n} = n + ik$ of (a) TaAs and (b) NbAs. The solid lines and dashed lines represent the experiment (Expt.) and DFT, respectively, with the smearing factor set to 0.2 eV. The blue and red curves are the real (Re) and imaginary (Im) components of the dielectric constant $\tilde{\varepsilon}$ and refractive index $\tilde{n}$. Subscripts 1 and 3 indicate directions along ordinary and extraordinary eigendirections.

## III. RESULTS AND DISCUSSION

### A. Anisotropic Dielectric Tensor: Experiments and DFT

The complex dielectric function $\tilde{\varepsilon}_r$ and refractive index $\tilde{n}$ obtained from both ellipsometry and DFT is shown in **Fig. 1** (see Fig. G1 for a smearing factor of 0.02 eV). The experimental and DFT results show a reasonable agreement in terms of the trends, magnitudes, and resonance frequencies; this level of agreement between DFT and experiments is considered acceptable [44,45] because DFT is a zero-temperature theory and has a well-known underestimation of the energy of excited states. The plasma edge determined in previous studies is below 0.1 eV [22,24]. Therefore, in the visible range, the dielectric function is dominated by interband transitions. Including the Drude behavior does improve the agreement between DFT and experiments below ~1.3eV. To capture the resonances from 1.2 eV to 6 eV, the Lorentz model is used to simulate phase and intensity changes of the reflected light in the ellipsometry data as shown below [26],

$$\tilde{\varepsilon} = \varepsilon_\infty + \sum_n^N \frac{A_n \Gamma_n E_n}{E_n^2 - E^2 - i \cdot E \Gamma_n} + \frac{A_{UV}}{E_{UV}^2 - E^2} + \frac{A_{IR}}{E^2} \tag{1}$$

where $A_n$, $\Gamma_n$, and $E_n$ are the resonance amplitude, broadening, and energies, respectively, of each oscillator labeled by $n$, and $\varepsilon_\infty$ represents the dielectric constant in the infinite frequency limit. Subscripts UV and IR are used as deep-ultraviolet and infrared poles, respectively, beyond the fitting spectrum. In TaAs, eight oscillators are captured along the crystallographic $a$ axis and seven oscillators are captured along the $c$ axis. The detailed fitting parameters for oscillators are shown in **Tables A.I** (for the ordinary component) and **A.II** (for the extraordinary component) for TaAs, and in **Tables A.III** (for the ordinary component) and **A.IV** (for the extraordinary component) for NbAs.

It is found that within 1.2 eV to 6 eV, $k$ decreases slower than $n$ and gradually exceeds the real component. Thus, with an increase in the photon energy, the reflectivities of TaAs and NbAs decrease, in agreement with previous literature [22], while the absorption coefficients show much less dispersion with energy. However, near the measured spectrum edges (below 1.2 eV and above 6 eV), the oscillators are not fully captured in the experiments, and only the tails of resonances can be observed. Therefore, the resonance energies of the first and last oscillators along each of the eigendirections are set slightly beyond the experimentally measured spectrum in order to simulate the resonance near the spectral edges. The broadening parameters, $\Gamma_n$ in **Eq. (1)** show a similar magnitude between experiments and DFT for a smearing factor of 0.2 eV, but are larger for the experimental resonances than for the DFT resonances when using a smearing factor 0.02 eV. The larger smearing factor obscures the details of several narrower DFT resonances (seen with a smearing factor of 0.02eV as shown in **Fig. G1**). We consider the larger smearing factor to be a finite temperature effect, and the smaller smearing factor to be a low-temperature effect.

Having shown the overall agreement between experiments and theory in **Fig. 1**, and given that the DFT can shed light on the electronic transitions at energies of relevance, we next focus our discussions below on the DFT resonances. The linear dielectric functions from DFT are fitted from 0 to 6 eV using **Eq. (1)**. The complete set of Expt. and DFT oscillators from 1.2-6 eV and 0-6 eV for TaAs are shown in **Fig. 2** (See **Fig. B1** for NbAs), and the detailed fitting parameters are shown in **Tables. AI-IV** (results of the smearing factor set to 0.02 eV can be found in **Appendix G**). The black curve shows the total $\varepsilon''$ and the colored curves represent the fitted oscillators. Nine oscillators are fitted along the *a* direction, and eight oscillators are fitted along the *c* direction from 0-6 eV. The resonances captured by experiments (Expt.) and density functional theory (DFT) show an excellent qualitative match and a reasonable quantitative match. In **Fig. 2(a)**, resonances in experiments (Expt.) and DFT at similar energies are mapped with the same colors; this mapping should be taken as approximate guides to the eye and not as a strict many-to-one mapping. Along the ordinary direction, experimental oscillators at 1.0, 1.8, 2.9, 3.7, and 4.1 eV can be approximately one-to-one mapped with DFT oscillators at 1.1, 2.0, 2.8, 3.5, and 4.1 eV, respectively. Along the extraordinary direction, experimental oscillators at 1.2, 1.8, 2.2, 2.8, 3.0, and 4.1 eV can be one-to-one mapped to the DFT resonances at 1.2, 1.9, 2.2, 2.9, 3.1, and 4.1 eV, as shown in **Fig. 2(b)**. A resonance dominates the dielectric constant at ~0.2 eV along the ordinary direction under a smearing factor of 0.02 eV (see **Fig. G2**, panel (b)), which can be associated with transitions from the lower Weyl bands (LWB) to the energy states above the upper Weyl bands (UWB) at WP2; this resonance is obscured with a larger smearing factor.

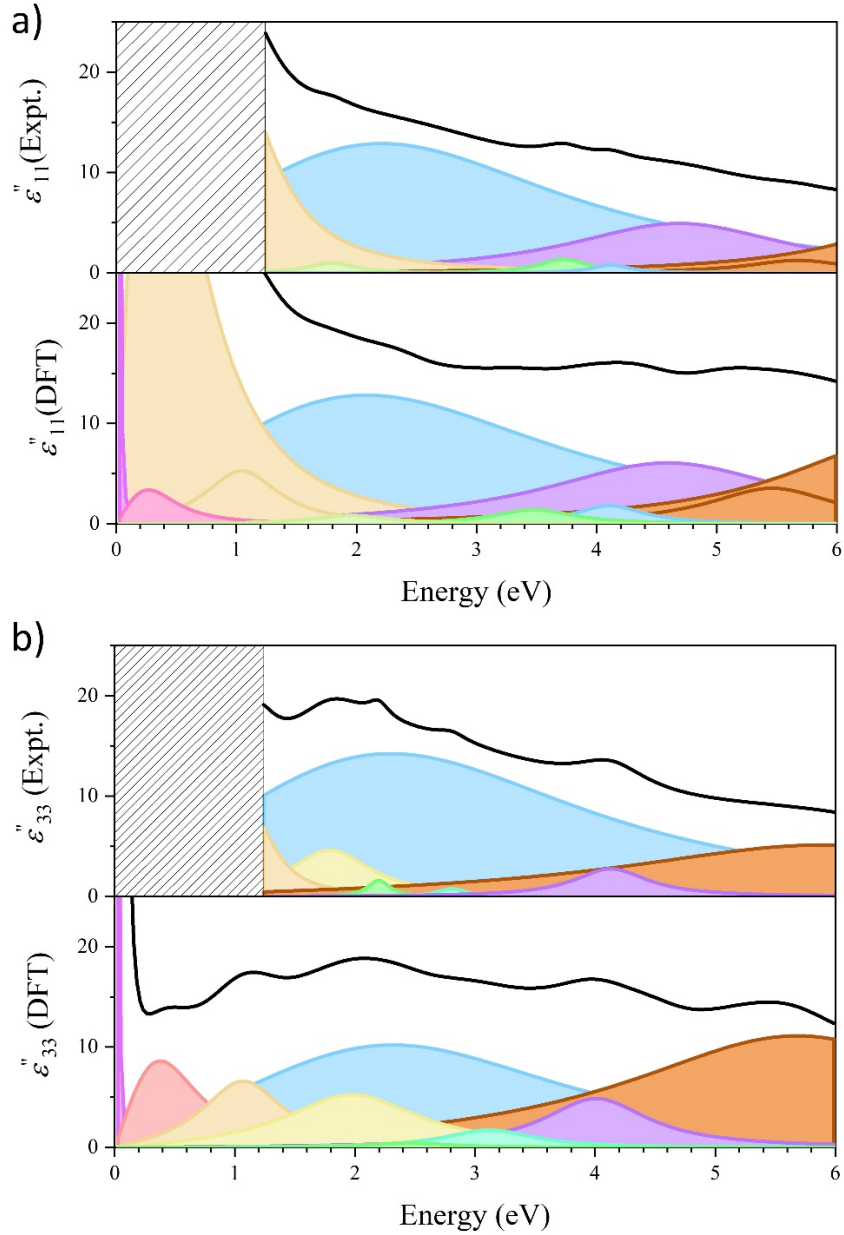

FIG 2. Detailed Lorentz oscillators for TaAs obtained from both experiments and DFT for a smearing factor of 0.2 eV. a) Ordinary imaginary dielectric function, $\varepsilon''_{11}$. The hatched area indicates the energy range beyond the experimentally measured spectrum. b) Extraordinary imaginary dielectric function, $\varepsilon''_{33}$. The black curve is the total imaginary dielectric function. Colored curves represent individual oscillators. In each subplot, similar colors of oscillators are approximate many-to-one mapping from theory to experiments; such mapping should not be strictly interpreted but only as a general guide to the eye as described in the main text. The resonance energies and linewidths are given in **Tables. AI-II**.

### B. Band Structure and Optical Transitions

Since the optical excitation occurs across the bands throughout the entire Brillouin zone (BZ), both transitions from the non-topological bands and the topological (Weyl) bands are reflected in the total dielectric spectrum, and care is needed in distinguishing their independent contributions. To understand the optical dielectric functions plotted in **Figs.1** and **2**, we carefully examine the electronic band structure and the optical transition matrix elements. With the confidence in the DFT calculated dielectric function based on its reasonable comparison with experiments (**Figure 1**), we next proceed to look carefully at their microscopic origins.

In **Fig. 3(a)**, the calculated DFT-SOC band structure, and the resulting density of states for TaAs are shown (the corresponding figure for the band structure and the density of states for NbAs are shown in **Appendix C, Figure C1**). Here, V1-8 and C1-8 are sequentially numbered valence and conduction bands, respectively, in the figures. Two different Weyl points (WP1 and WP2) were previously identified to be located along the $\Gamma - \Sigma$ and $\Sigma - \Sigma_1$ directions highlighted in the dashed box, in the absence of SOC (See **Fig. E1(b)** for high symmetry points). [2,46] When SOC is taken into account, the lines of Weyl nodes are further decomposed into 12 pairs of Weyl nodes, four pairs of WP1, and eight pairs of WP2. [47] Mirror symmetry protects each pair of Weyl points. The $k$-space momenta for WP1 and WP2 are calculated to be (0.015, 0.925, 0.004) and (0.505, 0.034, 0.313), where $k_x$, $k_y$ and $k_z$ are parallel to the crystallographic *a*, *b*, *c* axes, respectively. In the vicinity of the Fermi level, the states are mostly contributed by the hybridized Ta (*d*) orbital, while the states provided by As (*p*) are concentrated near the Z point. Deep within the valence band, the states are mostly As (*p*) character. The conduction band is mostly composed of Ta (*d*) orbitals. In the case of optical excitation in the infrared (IR) and visible regimes, optical excitations in TaAs are dominated by Ta (*d*) to Ta (*d*) and As (*p*) to Ta (*d*) transitions. In order to evaluate major optical transitions that contribute to the macroscopic resonances, the energy difference $\Delta\mathcal{E}_{na}$ is shown in **Fig. 3(b),** where $n$ and $a$ represent a pair of bands, labeled $n$ and $a$, as indicated in the legend next to the **panel (b)**. For example, the red plot in panel (b) represents the energy difference between the 3$^{rd}$ valence band, $a$=V3, and the first conduction band, $n$=C1, as indicated in panel (a). Similarly, other pairs of bands are represented by other colors as indicated.

To connect the macroscopic Lorentz resonances with the optical transitions in the band structure, we then evaluate theoretical resonances and *k*-dependent direct transitions. Here we focus on the oscillators below 3.5 eV, where most experimental optical characterizations are performed. The linear spectroscopic susceptibility is determined by electric dipole transition moments and their resonant and anti-resonant states expressed as, [48]

$$\chi_{ij}^{(1)}(\omega) = \frac{N}{\epsilon_0 \hbar} \Sigma_k \Sigma_{(n,a)} \left( \frac{\mu_{an}^i \mu_{na}^j}{(\omega_{na}-\omega)-i\gamma_{na}} + \frac{\mu_{na}^j \mu_{an}^i}{(\omega_{na}+\omega)+i\gamma_{na}} \right) \quad (2)$$

where $\omega_{na} = \omega_n - \omega_a = \Delta\mathcal{E}_{na}/\hbar$, between energy bands $n$ and $a$, and $\chi_{ij}^{(1)}$, $\mu$, $\omega$, $N$, $k$ and $\gamma$ are respectively the anisotropic susceptibility component (*i* and *j* are dummy variables representing directions), electric dipole transition moment, photon frequency, the number density of atoms, momentum space and decay rate related to the damping term. Here, $n$ and $a$ represent the excited and ground states. In DFT, the damping factor was taken to be $\gamma_{na} = 0.2$ eV, the same as the smearing factor as described earlier. The electric dipole transition moments, $\mu_{na}$ were calculated by DFT for every pair of bands, $n$ and $a$, at every point of the BZ in **Figure 3(a)**, i.e. $\mu_{na} \equiv f(n, a, k)$. To generate the total dielectric contribution for every energy shown on the right-hand-most panel in **Figure 3, panel (b)**, we perform the summation over every pair of bands, $(n,a)$, at every $k$-point along the high symmetry directions shown in panel (a) as indicated in **Eq. (2)**. The dominant contribution to the dielectric function at each energy, $\hbar\omega$, comes from the pairs of bands $n$, and $a$, such that $\Delta \mathcal{E}_{na} \approx \hbar\omega$; these dominant regions in the $k$-space are shown in panel (b) with light blue, yellow, and green shaded rectangular regions. For example, the shaded blue box shown in the $\Gamma - \Sigma$ path indicates a dominant contribution from this path to the $|\tilde{\varepsilon}_{11}|$ in the energy range from 2.2 to 4.4 eV; the Im($\tilde{\varepsilon}_{11}$) is shown as a blue curve on the right, to correspond to the blue rectangles on the left. Similarly, for example, the yellow rectangle boxes in the $\Gamma - X$ path indicates a dominant contribution from this path to the $|\tilde{\varepsilon}_{33}|$ in the energy range from 0.3-3 eV and 3.2-4.2 eV; the Im($\tilde{\varepsilon}_{33}$) is shown as a yellow curve on the right, to correspond to the yellow rectangles on the left. Finally, the green rectangles (overlap of blue and yellow rectangles) in the left correspond to regions which contribute dominantly to both $|\tilde{\varepsilon}_{11}|$ and $|\tilde{\varepsilon}_{33}|$.

Panels (c) and (d) in **Figure 3** plot the dominant contributions to $|\tilde{\varepsilon}_{11}|$ and $|\tilde{\varepsilon}_{33}|$ with a finer momentum and energy resolution of the contributing bands. For example, the peak in panel (c) at 0.2eV (the energy at the saddle point between the Weyl points) along the $\Sigma - \Sigma_1$ path indicates that this path contributes most strongly to the dielectric function $|\tilde{\varepsilon}_{11}|$. Further, this contribution comes from the pair of bands (V1, C2) and (V2, C1) as shown by the color-coding of the bar chart at this location (the color code is given in panel (b)); these are which are Ta (5$d$) to Ta (5$d$) transitions. The results show that the dielectric contribution to $|\varepsilon_{11}|$ at 0.2eV from the saddle point between the pair of Weyl points is dominated mostly by $\Sigma - \Sigma_1$ path, followed by $\Sigma - Z$, and $\Gamma - \Sigma$ paths. From 0.3 eV to 0.7 eV, the $k$ space that contributes to $|\tilde{\varepsilon}_{11}|$ changes to $\Sigma_1 - Z$. Within 0.7 eV to 1.5 eV, various bands in the $\Gamma - X$ dominate $|\tilde{\varepsilon}_{11}|$. The $Z - \Gamma$ starts to play a significant role from 1.5 eV to 2.3 eV because of a collection of resonant bands near the high symmetry Z point.

In the energy range from 2.3 to 3.5 eV, the predominant $k$ space contribution to the dielectric function shifts to $\Gamma - \Sigma$ because of the large joint density of states (JDOS). Along the crystallographic $c$ axis, the dielectric contribution to $|\tilde{\varepsilon}_{33}|$ at 0.2eV is dominated mostly by $\Sigma - \Sigma_1$ path, followed $\Gamma - X$ paths; the dominant contributing pairs of bands are (V1, C2), (V2, C1), and (V1, C1). At energy below 0.3 eV, the optical transitions occur mostly along $\Sigma - \Sigma_1$. Within 0.3 - 3.0 eV and 3.3 - 4.1 eV, $\Gamma - X$ is the primary $k$ space path that dominates the $|\tilde{\varepsilon}_{33}|$ as shown in the blue and green rectangles in **Fig. 3(b)**.

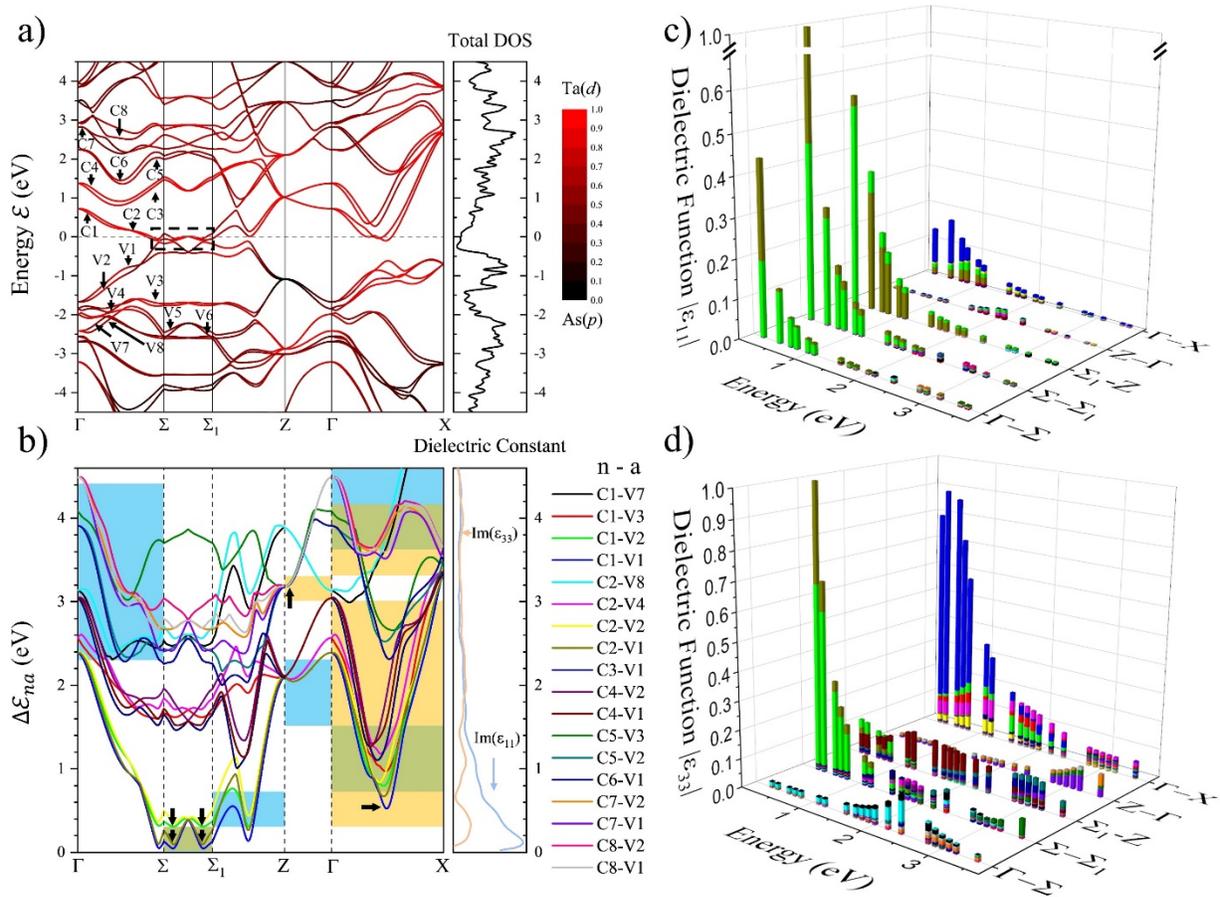

FIG 3. a) DFT-SOC band structure of TaAs. The color scale represents the relative band character of Ta ($d$) (red) vs. As($p$) (black) orbitals. Dashed rectangle highlights positions of Weyl points without SOC. b) The energy difference, $\Delta \mathcal{E}_{na}$, between selected pairs of bands, ($n,a$) listed on the right, contributing the most to the magnitude of the dielectric functions. The shaded areas highlight the paths and bands that contribute most to the dielectric function in that energy range. The blue, yellow, and green boxes correspond to pairs of bands contributing the most to $|\tilde{\varepsilon}_{11}|$, $|\tilde{\varepsilon}_{33}|$, and both, respectively. Arrows indicate Van Hove singularities that contribute to resonances. Normalized $k$-dependent contributions to the dielectric constants for the same pairs of bands as in (b) calculated along the c) ordinary and d) extraordinary directions, respectively using **Eq. (2)**. The color scale is the same as in panel (b). The smearing factor used for (b), (c) and (d) is 0.2 eV.

### C. Resonances close to observed optical second harmonic generation energies

Since $\chi^{(2)}_{333}$ is proportional to the product, $\tilde{\varepsilon}^2_{33}(\omega)\tilde{\varepsilon}_{33}(2\omega)$, and since resonances will enhance the linear dielectric permittivities, they will enhance the $\chi^{(2)}_{333}$ [49]. A large $\chi^{(2)}_{333}$ can be observed by a large transition dipole moment $\mu$ for a linear resonance, or a resonance close to either $\omega$ or $2\omega$ state [48]. Therefore, the direct optical transition with photon energy close to either $\omega$ or $2\omega$ states will enhance the SHG response. Optical second harmonic generation (SHG) measurements were previously performed at various incident fundamental energies such as 0.7 eV and 1.55 eV, and large $\chi^{(2)}_{333}$ coefficients were reported [25,26]. Experimental oscillators at 1.2, 1.8 and 3.0 eV, and DFT oscillators at 0.5, 1.2, 1.9, and 3.1 eV are close to the reported fundamental and SHG energies of interest, and thus likely resonantly enhancing the SHG. The $\Gamma - X$ dominates the overall $|\tilde{\varepsilon}_{33}|$ from 0.5 – 3 eV away from momentum space where Weyl points reside, indicating the trivial bands are the major resonance source of large $|\tilde{\varepsilon}_{33}|$ instead of Weyl-related states. Major optical transitions occur between V1-C1 along $\Gamma - X$ at 0.7 and 1.55 eV. It is worth noting that $Z - \Gamma$ exhibit large $|\tilde{\varepsilon}_{33}|$ between 3.0 – 3.3 eV due to resonating bands at high symmetry Z points which potentially resonant with the optical SHG experimentally observed at 3.1 eV in the previous literature [25,50]. In addition, there are contributions from Van Hove singularities and larger $\mu$'s contributing (see next section) to the SHG enhancement at 3.1eV energy. All of these resonances can thus enhance the observed SHG.

**D. Van Hove Singularities and Contributions from *k*-space in proximity to the Weyl Points**

A set of near-parallel flat bands or where there is a change in the slope of the $E(k)$ relation can produce Van Hove singularities (VHS) that give rise to large joint density of states (JDOS). [51–55] The critical points in the band structure exhibiting VHS have been previously reported at the saddle points connecting two Weyl nodes. [23,56] Along the extraordinary direction, the VHS is confirmed as highlighted by the dark arrows in **Fig. 3(b)** where $\nabla_k(\Delta\mathcal{E}_{na}) \approx 0$. [53,57] Here, $k$ is the lattice momentum and $\Delta\mathcal{E}_{na}$ is the energy difference between band $n$ and $a$. By identifying the $k$-space momenta, the VHS are located at $\Sigma - \Sigma_1$ below 0.4 eV, at $\Gamma - X$ near 0.5 eV and near high symmetry point $Z$ at 3.1 eV. The resonances contributed by VHS are confirmed by macroscopic oscillators at 0.5 and 3.1 eV. Moreover, the collection of bands near $Z$ at 3.1 eV further increases the JDOS. In addition, the magnitudes of dipolar transition matrix elements are larger near those critical points. As for the parallel band (V1, C1) as highlighted in the dark arrow near 0.7 eV, the magnitudes of dipolar transition matrix elements show the largest value and are up to near 100 times larger as compared with near-by momentum space. Near 1.55 eV, there is no VHS contributing along $\Gamma - X$ indicating the dielectric function at 1.55eV are majorly contributed by direct optical transitions. Similarly, near the high symmetry point $Z$, the magnitudes of dipolar transition matrix elements for the pair of bands (V1,C8), (V2, C7) and (V3, C5) dominate and are up to 85 times larger than the nearby momentum space; this enhancement directly contributes to the resonance at 3.1eV, and hence to the SHG detected at this energy. SHG is thus enhanced by direct optical transitions, VHS and large transition matrix elements.

Motivated by the resonances near the Weyl points, we then examine the contributions of the momentum space in the proximity of two different types of Weyl points to the dielectric functions. **Figure 4** shows the anisotropic dielectric functions contributed by a small volume of $k$-space (which is 0.0125% of the volume of the volume of the 1$^{st}$ BZ) around WP1 and WP2. For simplicity, in the discussion below, we will simply call these contributions as those "due to WP1 and WP2". The free carrier contributions are also included for both TaAs and NbAs to fulfill the metallic properties [22,58,59]. Though interband transitions are significant below 0.2 ~ 0.3 eV, the free carrier contribution to the dielectric function dominates. Therefore, below 0.5eV, the contributions from both WP1 and WP2 are below 25% of the total dielectric function in TaAs and NbAs.

Along the ordinary direction, both WP1 and WP2 show a similar contribution to $|\tilde{\varepsilon}_{11}|$ across the energy spectrum, and the magnitude of the dielectric functions tends to diminish after 0.2 eV. Along the extraordinary direction, the contribution to the $|\tilde{\varepsilon}_{33}|$ from WP2 tends to dominate over WP1, and the contribution is negligible above 1 eV. The DFT results in this study agree well with previous conclusions performed in the low energy range (0-0.2eV) using a linear band approximation [24]. Below 0.2 eV, the remaining dielectric function other than contributions from WP1 or 2 suggests they arise from dielectric resonances between trivial bands in both TaAs and NbAs. This anisotropic behavior in the lower energy range can be understood as different dispersions at two Weyl points using the approximate expression for the matrix elements for interband transitions, $\langle +\bm{k}|p_j|-\bm{k}\rangle^2 \propto \frac{1}{2}m^2(v_{F,+j}^2 + v_{F,-j}^2)$ [24]. Here, $v_F$, $k$ and $p$ are Fermi velocity, wavevector, and transition dipole operator, respectively. Subscript $j$ represents directions. Since the dispersion and Fermi velocity at WP2 along the polar axis is much larger than that of WP1, the dipolar transition matrix elements at WP2 is therefore larger, yielding a higher dielectric response along the polar axis. However, at higher energies, such as in the visible range where many optical measurements are typically performed, the contribution to dielectric function from both the Weyl points is negligible.

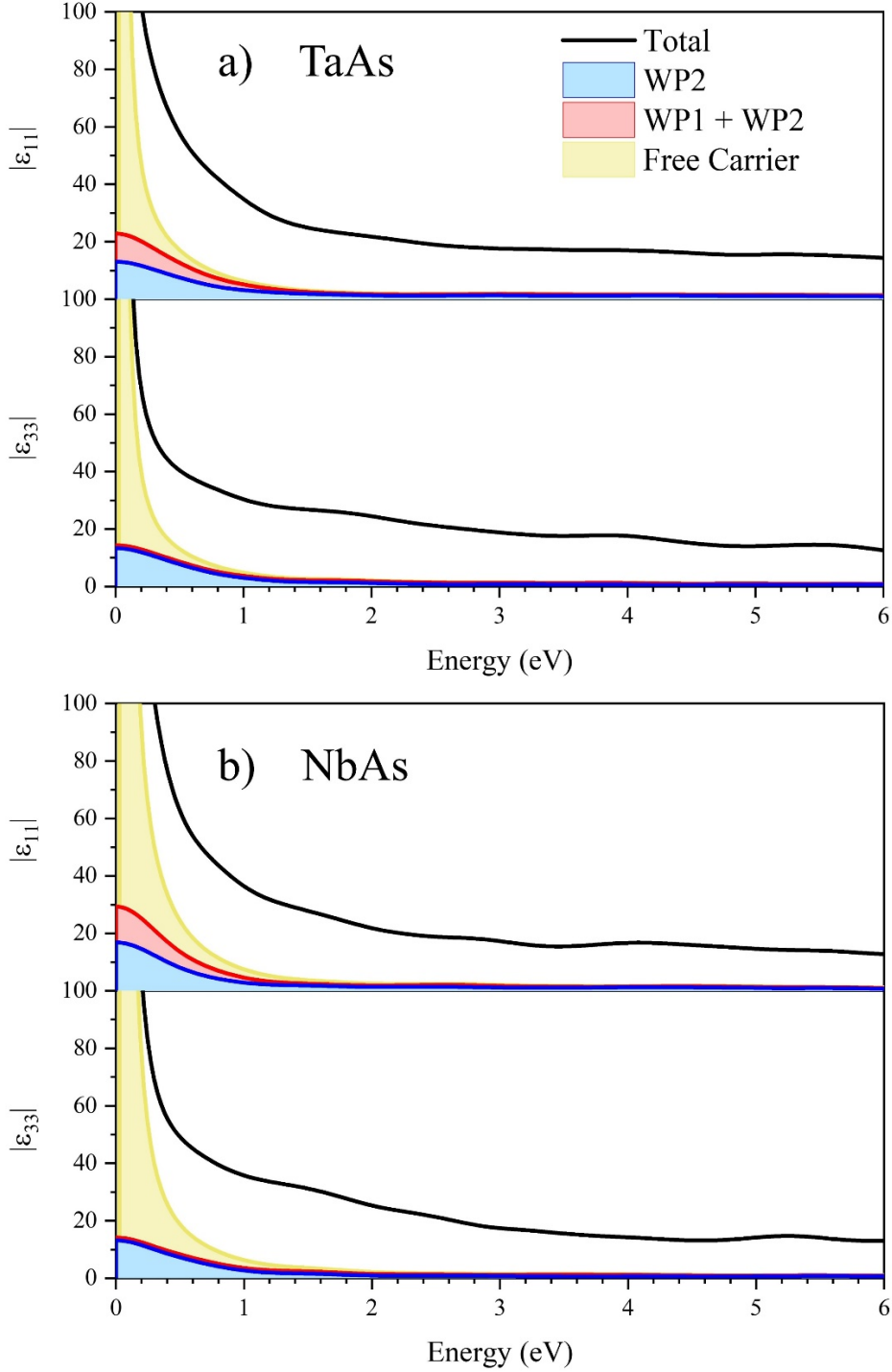

FIG 4. Stacked plots of the contributions of a small *k*-space (0.0125% of the volume of the 1$^{st}$ BZ) centered around the two types of Weyl points (WP1 and WP2), to the magnitudes of dielectric functions $|\tilde{\varepsilon}|$ for a smearing factor of 0.2 eV. a) Dielectric functions of TaAs. b) Dielectric

functions of NbAs. The upper panel represents the ordinary direction, and the bottom panel represents the extraordinary direction. The black (entire Brillouin zone), red (*k*-space centered at 8 WP1 + 16 WP2), blue (*k*-space centered 16 WP2) and yellow (free carrier) curves represent $|\tilde{\varepsilon}|$ contributions, as noted from different regions of the Brillouin zone.

## IV. CONCLUSIONS

TaAs family has attracted tremendous interest since the discovery of Weyl semimetals. In this work, for the first time, we summarize the anisotropic dielectric functions of both TaAs and NbAs across a broad spectrum combining ellipsometry and DFT. The magnitudes, trends, and resonance energy show an excellent match between experiments and theory. The energy-, momentum-, and band-resolved dielectric contributions clearly show resonance between trivial bands near the energies where giant SHG was reported earlier. Finally, for the first time, we analyze the contribution of a small volume of *k*-space (0.0125% of the volume of the 1$^{st}$ BZ) centered around the WP1 and the WP2 to the dielectric functions. It reveals that such contribution is significant only below 0.5 eV and is <25% of the total dielectric function; this contribution is insignificant above 1 eV, i.e., in the near-infrared, visible, and the ultraviolet range. These results should provide guidance for future optical probing studies of Weyl physics in these important materials.

## ACKNOWLEDGMENTS

R.Z. acknowledges useful discussions with Yoonsang Park, Hari Padmanabhan, Jingyang He, and Xinyu Liang. R. Z, L. M, Z. M, and V. G acknowledge support from the NSF MRSEC Center for Nanoscale Science, DMR-2011839. M.G. was supported by the DOE under grant no.

DE-SC0012375. J.M.R was supported by the Army Research Office (ARO) under grant no. W911NF-15-1-0017. Work at UCLA was supported by NSF DMREF program under the award NSF DMREF project DMREF-1629457.

**APPENDIX A: Parameters of Lorentz Oscillators from Ellipsometry and DFT**

Based on **Eq. 1**, parameters of oscillators in TaAs and NbAs are extrapolated and summarized in **Table A. I-IV**. The spectroscopic range of ellipsometry covers the oscillators from 1.2 - 6 eV (1000 - 200 nm). The oscillators from DFT (for a smearing factor of 0.2eV) are fitted from 0 - 6 eV to obtain a complete spectroscopic range and to compare with the experimental results.

TABLE AI. Parameters for the Lorentz oscillators given by **Eq. (1)** and the Drude model for the ordinary dielectric function $|\tilde{\varepsilon}_{11}|$ for TaAs. $\varepsilon^\infty = 1.8$ (Expt.) and 1.6 (DFT), $A_{UV} = 0.1$ (Expt.) and 4.0 (DFT), $E_{UV} = 6.6$ (Expt.) and 7.0 (DFT), $A_{IR} = 10.0$ (Expt.), $\sigma_D = 2 \times 10^6 \, ohm^{-1}m^{-1}$, $\tau_D = 3.3 \, ps$. [22]

| Oscillators (n) | Expt. $E_n$ ($\Gamma_n$) (eV) | Expt. $A_n$ (a.u.) | DFT $E_n$ ($\Gamma_n$) (eV) | DFT $A_n$ (a.u.) |
|---|---|---|---|---|
| 1 | 1.0(0.8) | 21.2 | 0.5(1.1) | 40.3 |
| 2 | 1.8(0.5) | 1.0 | 1.1(0.8) | 5.6 |
| 3 | 2.9(4.0) | 11.3 | 2.0(0.8) | 0.9 |
| 4 | 3.7(0.5) | 1.4 | 2.8(4.3) | 11.1 |
| 5 | 4.1(0.4) | 0.8 | 3.5(0.8) | 1.4 |
| 6 | 4.8(2.2) | 4.9 | 4.1(0.7) | 1.8 |
| 7 | 5.7(1.0) | 1.2 | 4.8(2.5) | 6.0 |
| 8 | 6.8(2.0) | 4.6 | 5.5(1.3) | 3.5 |
| 9 | - | - | 6.6(2.3) | 8.1 |

TABLE AII. Parameters for the Lorentz oscillators given by **Eq. (1)** and the Drude model for the extraordinary dielectric function $|\tilde{\varepsilon}_{33}|$ for TaAs. $\varepsilon^{\infty} = 1.7$ (Expt.) and 1.6 (DFT), $A_{UV} = 0.1$ (Expt.) and ~0 (DFT), $E_{UV} = 6.5$ (Expt.) and 6.7 (DFT), $A_{IR} = 6.6$ (Expt.), $\sigma_D = 2 \times 10^6\ ohm^{-1}m^{-1}$, $\tau_D = 3.3\ ps$. [22]

| Oscillators (n) | Expt. $E_n$ ($\Gamma_n$) (eV) | Expt. $A_n$ (a.u.) | DFT $E_n$ ($\Gamma_n$) (eV) | DFT $A_n$ (a.u.) |
|---|---|---|---|---|
| 1 | 1.2(0.5) | 8.0 | 0.5(0.9) | 7.0 |
| 2 | 1.8(0.8) | 4.6 | 1.2(0.9) | 6.2 |
| 3 | 2.2(0.2) | 1.6 | 1.9(0.8) | 2.5 |
| 4 | 2.8(0.3) | 0.8 | 2.2(0.5) | 1.1 |
| 5 | 3.0(4.4) | 12.4 | 2.9(0.3) | 0.04 |
| 6 | 4.1(0.8) | 2.8 | 3.1(4.1) | 12.2 |
| 7 | 6.2(4.1) | 5.0 | 4.1(0.9) | 3.5 |
| 8 | - | - | 6.1(3.6) | 10.0 |

TABLE AIII. Parameters for the Lorentz oscillators given by **Eq. (1)** and the Drude model for the ordinary dielectric function $|\tilde{\varepsilon}_{11}|$ for NbAs. $\varepsilon^{\infty} = 1.4$ (Expt.) and 0.01 (DFT), $A_{UV} = 1.4$ (Expt.) and 8.3 (DFT), $E_{UV} = 6.7$ (Expt.) and 7.0 (DFT), $A_{IR} = 12.2$ (Expt.), $\sigma_D = 1 \times 10^7\ ohm^{-1}m^{-1}$, $\tau_D = 7.1\ ps$. [58,59]

| Oscillators (n) | Expt. $E_n$ ($\Gamma_n$) (eV) | Expt. $A_n$ (a.u.) | DFT $E_n$ ($\Gamma_n$) (eV) | DFT $A_n$ (a.u.) |
|---|---|---|---|---|
| 1 | 1.1(0.2) | 11.3 | 0.3(0.4) | 35.3 |
| 2 | 1.7(2.4) | 17.3 | 0.9(1.2) | 17.4 |
| 3 | 3.0(0.9) | 3.4 | 1.9(2.1) | 11.7 |
| 4 | 4.4(1.5) | 2.6 | 2.9(1.0) | 3.3 |
| 5 | 5.9(5.6) | 5.1 | 4.3(1.3) | 3.2 |
| 6 | - | - | 6.0(5.8) | 11.6 |

TABLE AIV. Parameters for the Lorentz oscillators given by **Eq. (1)** and the Drude model for the extraordinary dielectric function $|\tilde{\varepsilon}_{33}|$ for NbAs. $\varepsilon^\infty = 2.3$ (Expt.) and 1.0 (DFT), $A_{UV} = 3.4$ (Expt.) and 64.9 (DFT), $E_{UV} = 6.7$ (Expt.) and 7 (DFT), $A_{IR} = 8.3$ (Expt.), $\sigma_D = 1 \times 10^7 \ ohm^{-1}m^{-1}$, $\tau_D = 7.1 \ ps$. [58,59]

| Oscillators (n) | Expt. $E_n$ ($\Gamma_n$) (eV) | Expt. $A_n$ (a.u.) | DFT $E_n$ ($\Gamma_n$) (eV) | DFT $A_n$ (a.u.) |
|---|---|---|---|---|
| 1 | 1.5(1.2) | 20.1 | 0.3(0.2) | 2.0 |
| 2 | 1.9(0.2) | 1.9 | 0.8(0.8) | 8.5 |
| 3 | 2.6(1.5) | 10.2 | 1.6(1.4) | 12.0 |
| 4 | 3.5(0.7) | 3.0 | 1.8(0.5) | 1.0 |
| 5 | 4.6(2.8) | 4.7 | 2.6(2.0) | 11.0 |
| 6 | 5.5(0.8) | 3.5 | 3.5(1.1) | 2.1 |
| 7 | 6.4(1.0) | 2.9 | 4.5(3.0) | 6.8 |
| 8 | - | - | 5.5(1.5) | 8.1 |
| 9 | - | - | 6.3(1.3) | 1.8 |

### APPENDIX B: ANISOTROPIC DIELECTRIC TENSOR IN NbAs: EXPERIMENTS AND DFT

**Fig. B1** shows the complete set of Expt. and DFT oscillators from 1.2-6 eV and 0-6 eV in NbAs. The black curve shows the total $\varepsilon''$ and the colored curves represent each oscillator. Like TaAs, six oscillators are fitted along the $a$ direction, and nine oscillators are fitted along the $c$ direction for DFT from 0-6 eV. Strong resonances can be identified in the low energy range, which can be associated with the interband transitions near the Weyl points. [60] **Figure B1(a)** shows Expt. and DFT resonances in NbAs along the ordinary direction, and similar resonances are mapped with the same colors purely as an approximate guide to the eye. Experimental oscillators at 1.1, 1.7, 3.0, and 4.4 eV can be approximately mapped one-to-one with DFT oscillators at 0.9, 1.9, 2.9, and 4.3 eV. **Figure B1(b)** shows oscillators along the extraordinary direction. Resonances (Expt.) at 1.5, 1.9, 2.6, 3.5, and 4.6 eV can be relatively mapped one-to-one with DFT oscillators at 1.6, 1.8, 2.6, 3.5, and 4.5 eV, respectively.

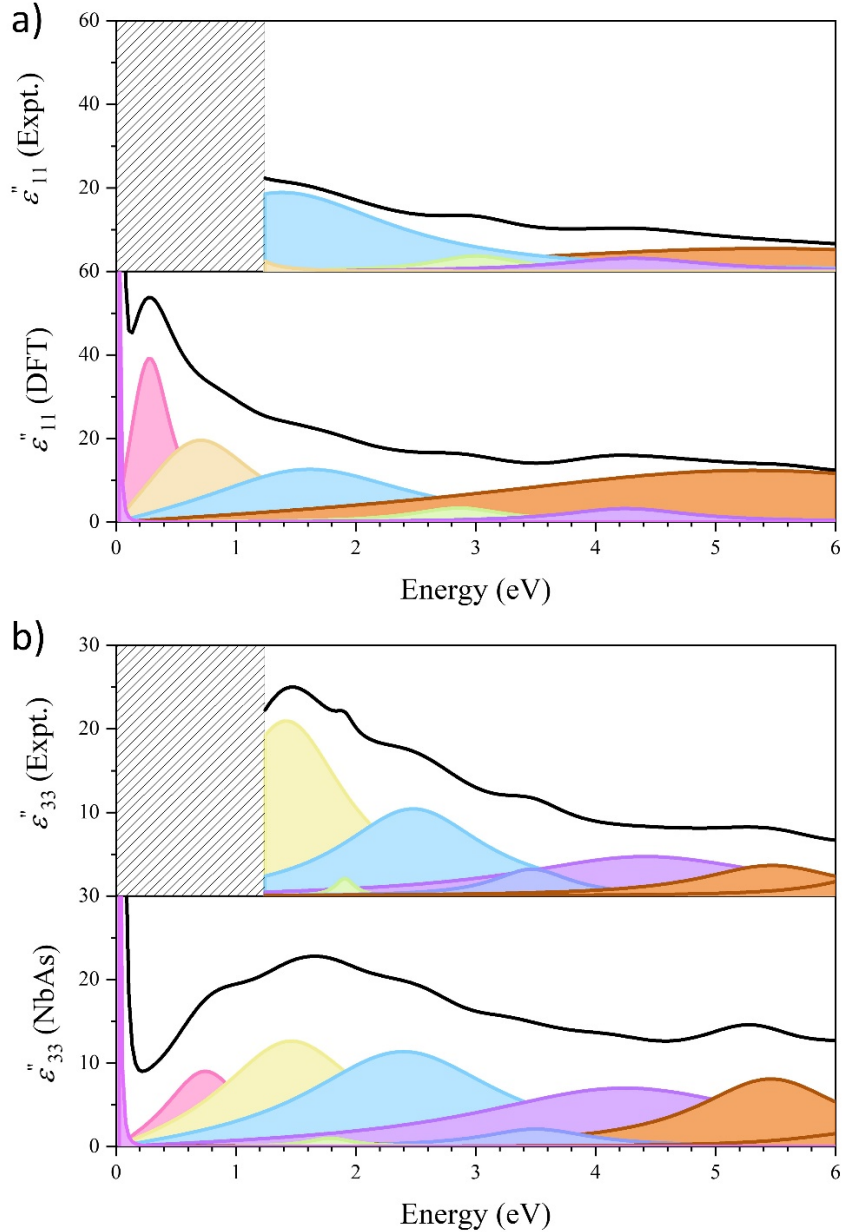

FIG B1. Detailed Lorentz oscillators for NbAs obtained from both experiments and DFT for a smearing factor of 0.2 eV. a) Ordinary imaginary dielectric function, $\varepsilon_{11}''$. The hatched area

indicates the energy range beyond the experimentally measured spectrum. b) Extraordinary imaginary dielectric function, $\varepsilon_{33}''$. The black curve is the total imaginary dielectric function. Colored resonance fits represent individual oscillators. In each subplot, similar colors of oscillators are approximately many-to-one mapping from theory to experiments; they are only to be taken as a guide to the eye. The resonance energies and linewidths are given in **Tables. AIII-IV.**

## APPENDIX C: BAND STRUCTURE, BAND- AND MOMENTUM- RESOLVED DIELECTRIC RESPONSE IN NbAs

In **Fig. C1(a)**, the calculated DFT-SOC band structure, and the resulting density of states for NbAs are shown. V1-10 and C1-8 are sequentially numbered valence and conduction bands. The band splitting by SOC in NbAs is smaller compared with TaAs due to the smaller atomic number of Nb [61]. The *k*-space momenta for WP1 and WP2 are calculated to be (0.891, 0.005, 0) and (0.012, 0.503, 0.306), respectively. Near the Fermi level, the states are mostly contributed by the hybridized Nb (*d*) orbitals. Deep within the valence state, the states are mostly dominated by the As (*p*) orbitals. In the case of optical excitation in the infrared (IR) and visible regimes, optical excitations in the system are dominated by Nb (*d*) to Nb (*d*) and As (*p*) to Nb (*d*) transitions.

**Fig. C1(b)** shows the energy difference $\Delta \mathcal{E}_{na}$ between pairs of bands contributing most to the dielectric function using **Eq. (2)**. The colored rectangles show the momentum space and pairs of bands that predominantly contribute to a certain dielectric constant value using the same scheme presented in **Fig. 3(b)**. In the **Figs. C1(c)** and **(d)**, the dominant pairs of bands that contribute to $|\tilde{\varepsilon}_{11}|$ and $|\tilde{\varepsilon}_{33}|$ are shown, with a finer momentum and energy-resolved picture.

Along the ordinary direction, from 0 to 0.5 eV, the optical transitions mainly arise from $\Sigma_1 - Z$. Further, this contribution mainly comes from the pairs of bands (V1, C2) and (V2, C1), which mostly originate from Nb (*d*) to Nb (*d*) and As (*p*) to Nb (*d*) transitions. From 0.5 eV to 1 eV, the $k$ space that contributes predominantly to $|\tilde{\varepsilon}_{11}|$ changes to $\Gamma - X$. $Z - \Gamma$ becomes significant from 1 eV to 1.9 eV because of a cluster of resonant bands near the high symmetry Z point, which corresponds to the macroscopic resonance (DFT) at 1.6 eV. In the energy range from 1.9 to 4.2 eV, the predominant k space shifts to $\Gamma - \Sigma$. Similarly, along the extraordinary direction of NbAs, the optical transitions below 0.2 eV mainly arise from $\Sigma - \Sigma_1$ near the Weyl nodes without SOC [46]. In this energy range, the primary optical transitions are identified to be (V2, C1), (V1, C2), and (V1, C1). In the 0.2 eV to 4.5 eV range, $\Gamma - X$ is the primary k space that dominates the $|\tilde{\varepsilon}_{33}|$.

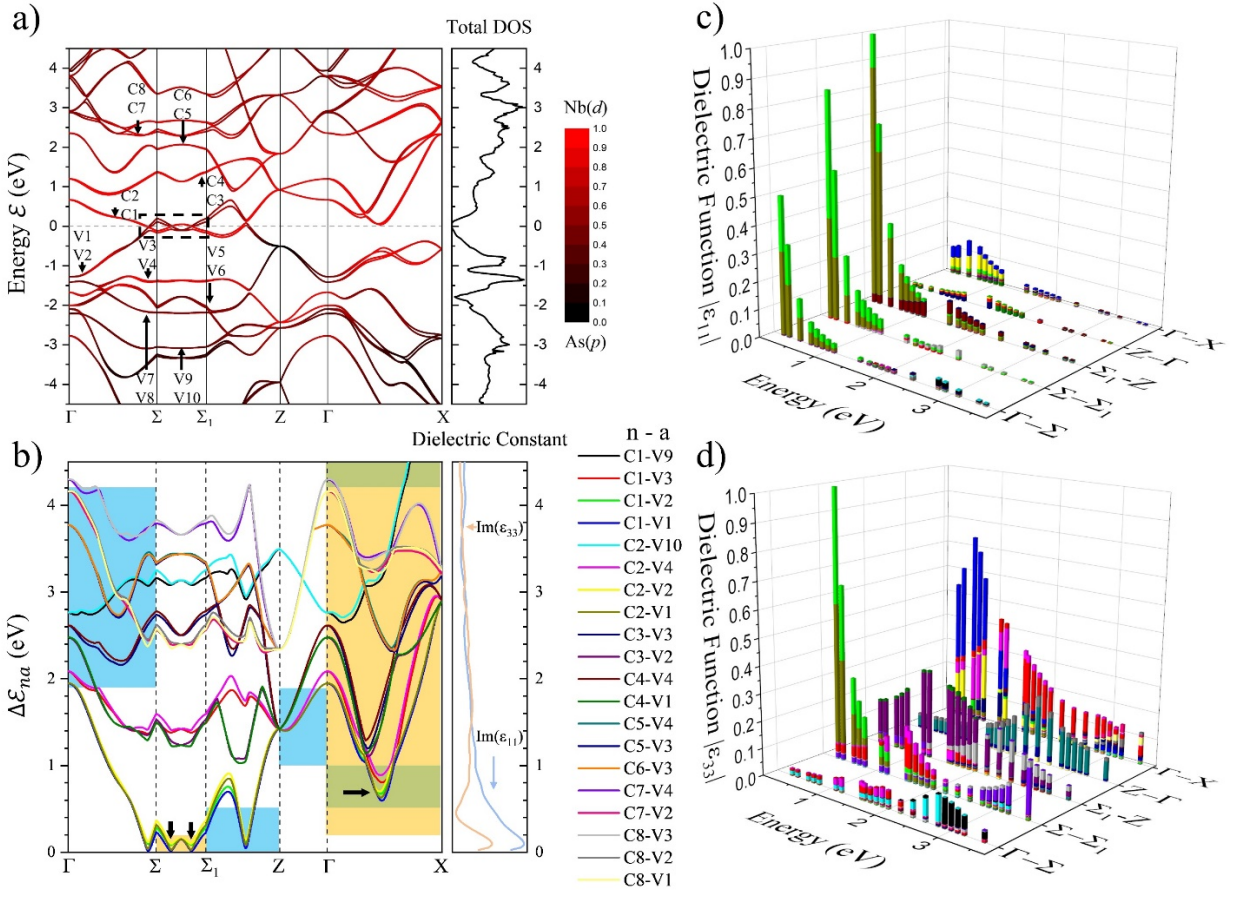

FIG C1. a) DFT-SOC band structure of NbAs. The color scale represents the relative band character of Ta (*d*) (red) vs. As(*p*) (black) orbitals. Dashed rectangle highlights positions of Weyl points without SOC. b) The energy difference, $\Delta\mathcal{E}_{na}$, between selected pairs of bands, $(n,a)$ listed on the right, contributing the most to the magnitude of the dielectric functions. The shaded areas highlight the paths and bands that contribute most to the dielectric function in that energy range. The blue, yellow, and green boxes correspond to pairs of bands contributing the most to $|\tilde{\varepsilon}_{11}|$, $|\tilde{\varepsilon}_{33}|$, and both, respectively. Arrows indicate Van Hove singularities that contribute to resonances. Normalized *k*-dependent contributions to the dielectric constants for the same pairs of bands as in (b) calculated along the c) ordinary and d) extraordinary directions, respectively using **Eq. (2)**. The color scale is the same as in panel (b). The smearing factor used for (b), (c) and (d) is 0.2 eV.

# APPENDIX D: RESONANCES NEAR OBSERVED SECOND HARMONIC GENERATION ENERGIES AND VAN HOVE SINGULARITIES IN NbAs

By examining resonances near 0.7, 1.55, and 3.1 eV along the extraordinary direction, experimental oscillators at 1.5 eV and 2.6 eV are close to the energies of interest. DFT oscillators at 0.8, 1.6, and 2.6 eV are near the reported energies. The $\Gamma - X$ dominates the overall $|\tilde{\varepsilon}_{33}|$ at energies of interest, away from the Weyl points. The primary optical transition occurs between V1-C1 along $\Gamma - X$ at 0.7 eV. At 1.55 eV, the major optical transitions are V4-C2 and V3-C1. Therefore, the enhancement at 1.55 eV is significantly contributed by the $\Gamma - X$ path. At 3.1 eV, more bands are involved in the optical transitions, while V2-C7 and V4-C2 contribute more to the dielectric function than the contribution from other transitions.

Similar to TaAs, VHS contributes to resonances along the extraordinary direction, as highlighted by the black arrows in **Fig. C1(b).** Below 0.2 eV, VHS can be observed at $\Sigma - \Sigma_1$. Near 0.7 eV, where SHG measurements were performed in this family, VHS at $\Gamma - X$ is expected to enhance the SHG response. Moreover, the dipole moments $\mu$ at the energy location of the dark arrows are larger than the rest of the *k*-space. Therefore, direct optical transitions, VHS, and large dipolar transition matrix elements can enhance $|\tilde{\varepsilon}_{33}|$ and further increase the response near the fundamental and SHG energies.

# APPENDIX E: SYNTHESIS AND STRUCTURAL CHARACTERIZATION

Single crystals of TaAs and NbAs were grown by chemical vapor transport. The stoichiometric ratio of Ta (Nb) and As powder was thoroughly mixed, pressed into a pellet, and sealed under a third atmospheric pressure of Ar in a quartz tube for the solid-state reaction. The tube was heated to 600°C at 50°C per hour, stayed for 10 hours, then heated to 1100°C at 50°C per hour, and finally dwelled for another 10 hours before water quenching. Three grams of such precursors are loaded with iodine pieces in a quartz tube. The density of the iodine was 18 mg/cm³. The quartz ampule was sealed under vacuum and put in a three-zone furnace for a four-week-long chemical vapor transport growth. The temperature was set to be 1050°C (950°C) at the source end and 950°C (850°C) at the sink end for TaAs (NbAs). Powder X-ray data were collected using a PANalytical Empyrean diffractometer with Cu-Kα radiation. As-grown surfaces of (020) can be identified and confirmed with the X-ray diffraction, which was polished to a larger size using 1-micrometer sandpapers and later used for the spectroscopic study.

The crystal structure and Brillion zone of TaAs and NbAs are shown in **Fig. E1**. The yellow represents As atoms, and grey represents Ta or Nb atoms. The Brillion zone and high symmetry points are labeled in **Fig. E1(b)**.

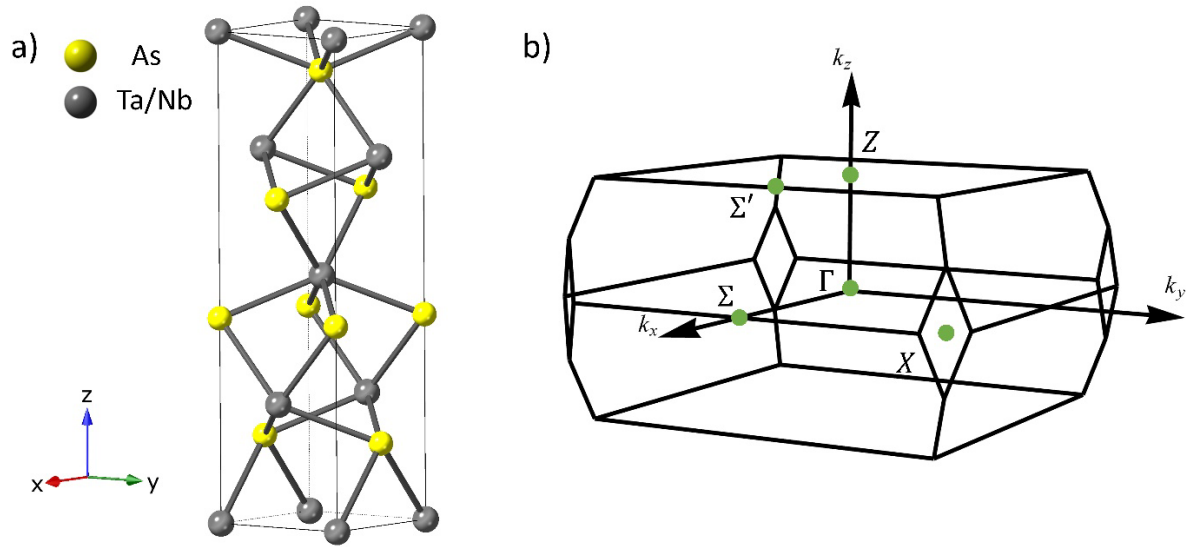

FIG E1. a) Crystal structure of TaAs and NbAs. b) The Brillion zone and high symmetry points of TaAs and NbAs.

Single crystalline TaAs and NbAs are confirmed using $\theta$-$2\theta$ X-ray diffraction (XRD), as shown in **Fig. E2**. The out-of-plane direction is confirmed to be (020).

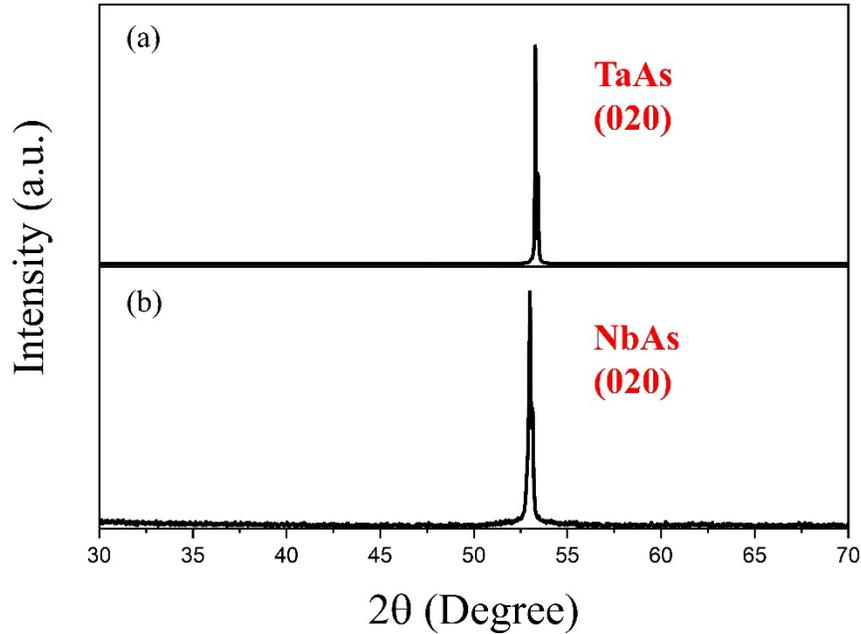

FIG E2. XRD of TaAs (020) and NbAs (020) plane. a) Experimental XRD of the TaAs (020) peak. b) Experimental XRD of NbAs (020) peak.

The crystals' orientations are determined by combining Laue X-ray backscatter diffraction and the electron backscatter diffraction (EBSD). The raw crystals are shown in the upper panel of **Fig. E3**. By orientating the crystal axis parallel to the *x-y-z* coordinates (*z* as out-of-plane direction) in the EBSD system, the orientation maps are obtained with (020) intensity peaks along the *z*-direction, confirming the previous XRD pattern in **Fig. E2**. The [001] direction peaks along *y* for TaAs, and *x* for NbAs. The orientations are identified and aligned correctly for further anisotropic characterization.

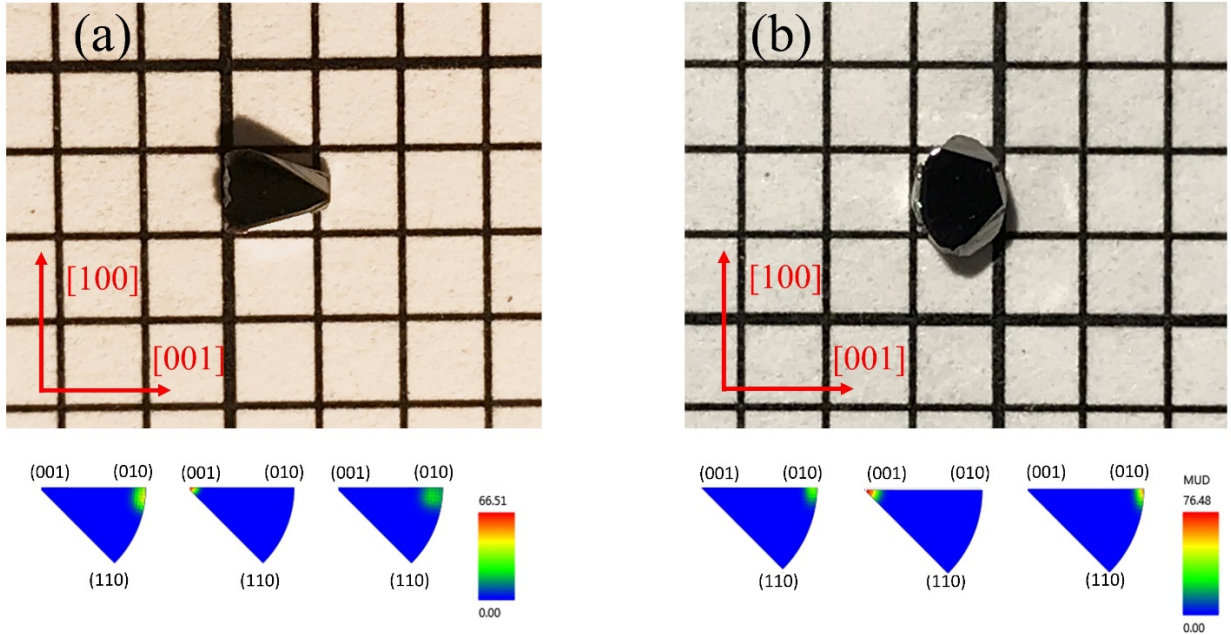

FIG E3. Optical images and inverse pole figures of synthesized a) TaAs and b) NbAs. The upper panel is the optical image of the studied crystals. The red arrows indicate crystallographic orientations. The size of the square boxes in the background is 2mm x 2 mm. The bottom panel is the inverse pole figures obtained from the EBSD. The out-of-plane direction is along [010].

## APPENDIX F: TRANSPORT CHARACTERIZATION

To characterize the quality of studied crystals, electrical and magnetoelectrical transport measurements were conducted with a standard four-probe method in a Quantum Design PPMS. The electric current was applied within the plane, while the magnetic field was along the out-of-plane direction for both TaAs and NbAs. To eliminate the non-symmetricity of the contacts, the longitudinal resistivity ($\rho_{xx}$) and Hall resistivity ($\rho_{xy}$) are symmetrized using the formulas $\rho_{xx}(H) = [\rho_{xx}(H) + \rho_{xx}(-H)]/2$ and $\rho_{xy}(H) = [\rho_{xy}(H) - \rho_{xy}(-H)]/2$, respectively, where $H$ is the magnetic field.

The magnetic field dependent longitudinal resistivities of TaAs and NbAs are shown in **Fig. F1(a)**. In the low-temperature range (<100K), Hall resistivity (**Fig. F1(b)** for TaAs; **Fig. F1(c)** for NbAs) shows large negative values, nonlinear field dependences, and SdH oscillations, indicating both electrons and holes are involved in transport and electrons make dominant contributions to transport. Shubnikov–de Haas (SdH) oscillations are clearly observed at 2K in both field sweep measurements of longitudinal and Hall resistivity, shown in **Figs. F1(a), (b) and (c)**.

For TaAs, we have estimated the carrier densities and mobilities of low temperatures through the fits of longitudinal and Hall resistivity at 10K by a two-band model: $n_e(10K) = 7.12 \times 10^{23}\ m^{-3}$ and $n_h(10K) = 5.56 \times 10^{23}\ m^{-3}$, and mobilities $\mu_e(10K) = 6.04\ m^2V^{-1}s^{-1}$ and $\mu_h(10K) = 2.54\ m^2V^{-1}s^{-1}$. All these results above are comparable with previously reported data of TaAs single crystals [36–38]. In NbAs, strong SdH oscillation is also observed. Compared with TaAs, the Hall signal of NbAs at 2K is more linear and can be analyzed by a single-band model. Similar to the previous report [39], the electron carrier density and electron mobility of the NbAs crystal is $n_e(10K) = 2.7 \times 10^{25}\ m^{-3}$ and $\mu_e(10K) = 83.5\ m^2V^{-1}s^{-1}$, respectively.

The fitting of TaAs transport data is presented in **Fig. F1(d)**. The fitting of the low-temperature magnetoelectrical transport data is based on a simplified two-band model, where only one electron band and one hole band are considered. The longitudinal resistivity ($\rho_{xx}$) and transverse resistivity ($\rho_{xy}$) can be described as [62]

$$\rho_{xx} = \frac{(n_e\mu_e+n_h\mu_h)+(n_e\mu_e\mu_h^2+n_h\mu_h\mu_e^2)B^2}{(n_e\mu_e+n_h\mu_h)^2+\mu_h^2\mu_e^2(n_h-n_e)^2B^2} \cdot \frac{1}{e} \tag{F1}$$

$$\rho_{xy} = \frac{(n_h\mu_h^2-n_e\mu_e^2)+\mu_h^2\mu_e^2(n_h-n_e)B^2}{(n_e\mu_e+n_h\mu_h)^2+\mu_h^2\mu_e^2(n_h-n_e)^2B^2} \cdot \frac{B}{e} \tag{F2}$$

where $n_e$ ($n_h$) and $\mu_e$ ($\mu_h$) are the density and mobility of the electron (hole) band, respectively. To constrain the freedom of the fitting, in this work, the relationship between $n_e$ and $n_h$ is estimated by the linear fitting of the $\rho_{xy}$ in the high-field range as $\rho_{xy} = \frac{1}{n_h - n_e} \cdot \frac{B}{e}$ [37]. The fitting of NbAs transport data is presented in **Fig. F1(e)**. The fitting of the low-temperature magnetoelectrical transport data of NbAs is based on a single band approximation. The carrier density is described by $n = \frac{1}{eR_H}$, where $R_H$ is the Hall coefficient. The mobility is further given by $\mu = \frac{1}{en\rho_{xx}(0T)}$, where $\rho_{xx}(0T)$ is the longitudinal resistivity at zero magnetic field.

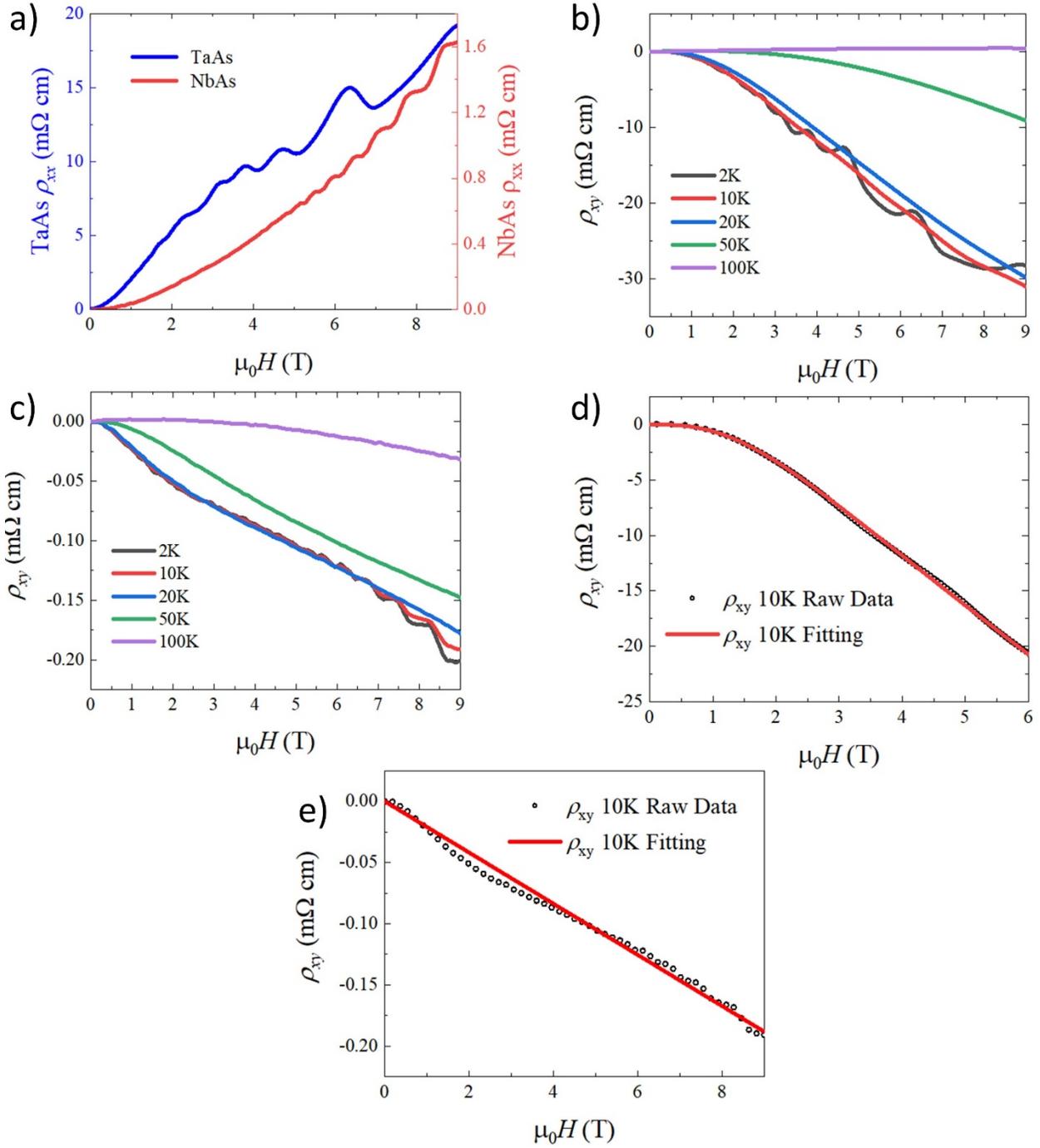

FIG F1. Transport Properties of TaAs and NbAs. (a) The magnetic field dependent longitudinal resistivity of TaAs and NbAs at 2K. The Hall resistivity versus magnetic fields and 50 to 2K in TaAs (b) and NbAs (c). (d) The two-band model fitting of Hall resistivity for TaAs at 10K. (e) The single-band model fitting of Hall resistivity of NbAs at 10K.

# APPENDIX G: A CASE STUDY FOR A SMEARING FACTOR OF 0.02 eV

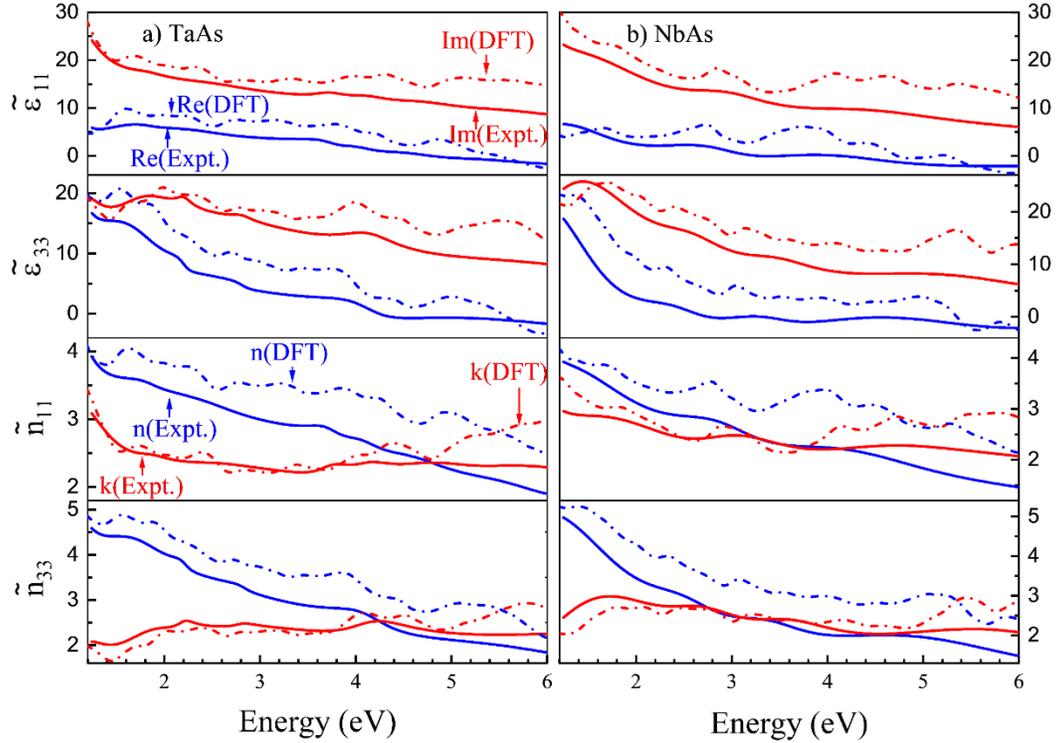

FIG G1. The complex dielectric functions $\tilde{\varepsilon}_r$ and refractive indices $\tilde{n} = n + ik$ of (a) TaAs and (b) NbAs. The solid line and dashed line represent the experiment (Expt.) and DFT with the smearing factor 0.02 eV, respectively. The blue and red curves are the real (Re) and imaginary (Im) components of the dielectric constant $\tilde{\varepsilon}$ and refractive index $\tilde{n}$. Subscripts 1 and 3 indicate directions along ordinary and extraordinary Eigendirections.

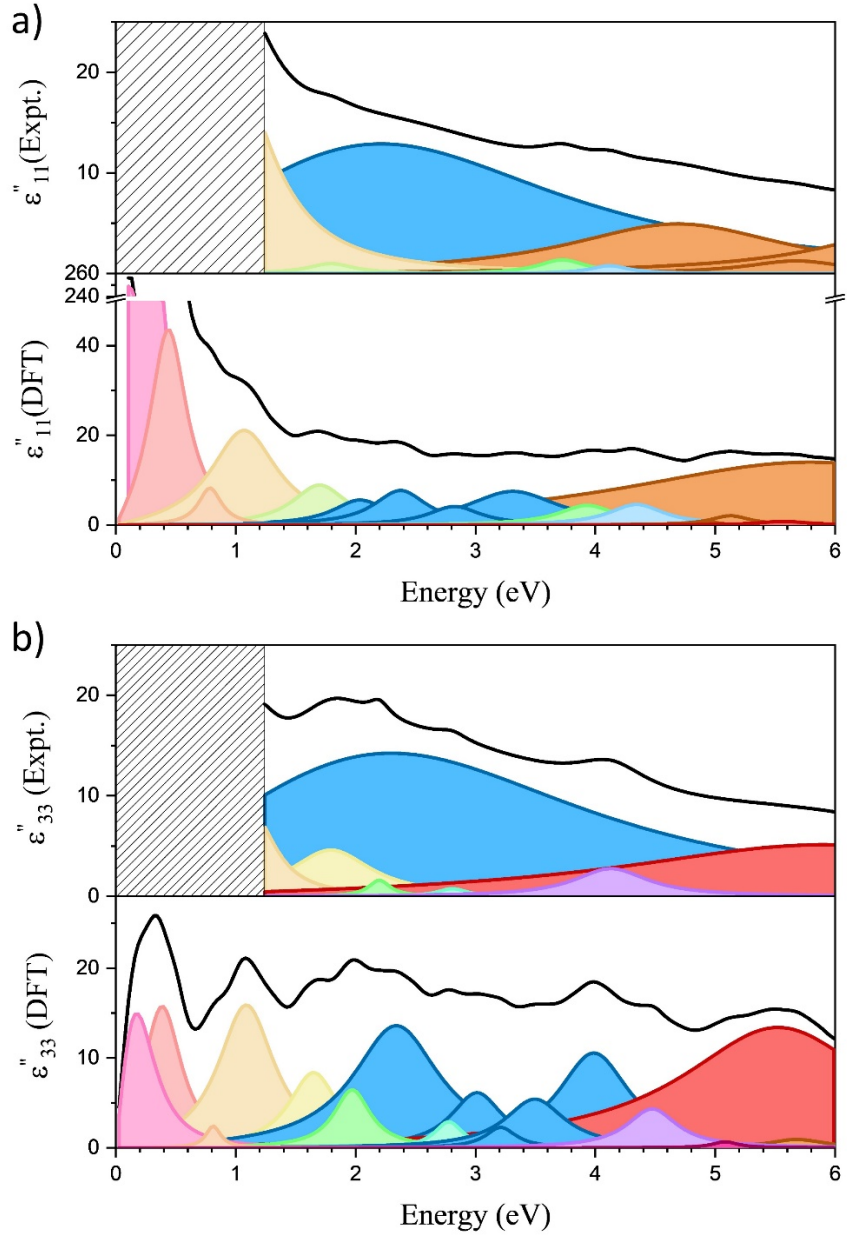

FIG G2. Detailed Lorentz oscillators for TaAs obtained from both experiments and DFT for a smearing factor of 0.02 eV. a) Ordinary imaginary dielectric function, $\varepsilon''_{11}$. The hatched area indicates the energy range beyond the experimentally measured spectrum. b) Extraordinary imaginary dielectric function, $\varepsilon''_{33}$. The black curve is the total imaginary dielectric function. Colored curves represent individual oscillators. In each subplot, similar colors of oscillators are approximate many-to-one mapping from theory to experiments; such mapping should not be strictly interpreted but only as a general guide to the eye as described in the main text. The resonance energies and linewidths are given in **Tables. GI-II**.

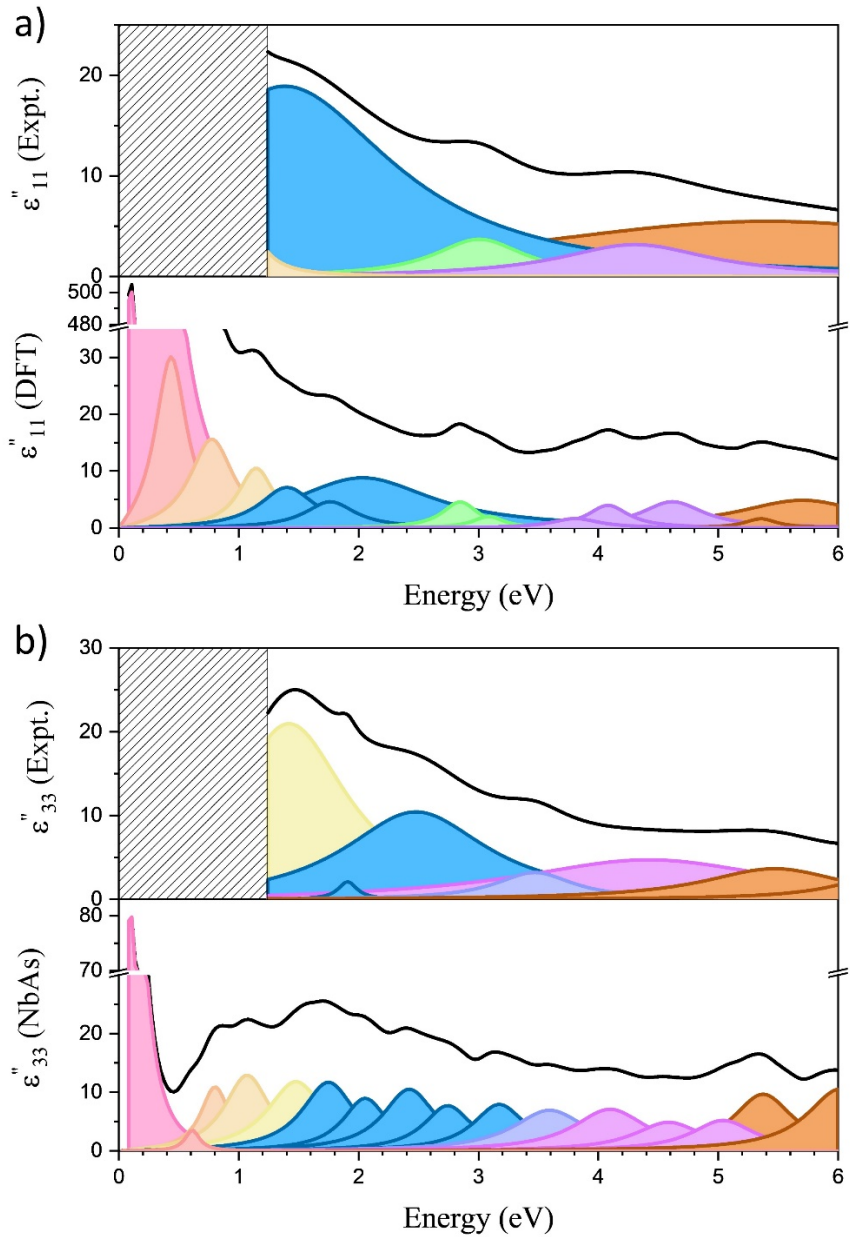

FIG G3. Detailed Lorentz oscillators for NbAs obtained from both experiments and DFT for a smearing factor of 0.02 eV. a) Ordinary imaginary dielectric function, $\varepsilon''_{11}$. The hatched area

indicates the energy range beyond the experimentally measured spectrum. b) Extraordinary imaginary dielectric function, $\varepsilon_{33}''$. The black curve is the total imaginary dielectric function. Colored resonance fits represent individual oscillators. In each subplot, similar colors of oscillators are approximately many-to-one mapping from theory to experiments; they are only to be taken as a guide to the eye. The resonance energies and linewidths are given in **Tables. GIII-IV.**

Based on **Eq. 1**, parameters of oscillators in TaAs and NbAs are extrapolated and summarized in **Table. GI-IV** for a smearing factor 0.02 eV. The spectroscopic range of ellipsometry covers the oscillators from 1.2 - 6 eV (1000 - 200 nm). The oscillators from DFT are fitted from 0 - 6 eV to obtain a complete spectroscopic range and to compare with the experimental results.

TABLE GI. Parameters for the Lorentz oscillators given by **Eq. (1)** for the ordinary dielectric function $|\tilde{\varepsilon}_{11}|$ for TaAs. $\varepsilon^\infty = 1.8$ (Expt.) and 1.8 (DFT), $A_{UV} = 0.1$ (Expt). and 34.5 (DFT), $E_{UV} = 6.6$ (Expt.) and 7.0 (DFT), $A_{IR} = 10.0$ (Expt.).

| Oscillators (n) | Expt. $E_n$ ($\Gamma_n$) (eV) | Expt. $A_n$ (a.u.) | DFT $E_n$ ($\Gamma_n$) (eV) | DFT $A_n$ (a.u.) |
|---|---|---|---|---|
| 1 | 1.0(0.8) | 21.2 | 0.2(0.3) | 200.7 |
| 2 | 1.8(0.5) | 1.0 | 0.5(0.4) | 41.9 |
| 3 | 2.9(4.0) | 11.3 | 0.8(0.2) | 8.2 |
| 4 | 3.7(0.5) | 1.4 | 1.1(0.7) | 20.9 |
| 5 | 4.1(0.4) | 0.8 | 1.7(0.6) | 9.0 |
| 6 | 4.8(2.2) | 4.9 | 2.1(0.6) | 6.1 |
| 7 | 5.7(1.0) | 1.2 | 2.4(0.5) | 7.9 |
| 8 | 6.8(2.0) | 4.6 | 2.8(0.5) | 4.3 |
| 9 | - | - | 3.4(1.0) | 9.3 |
| 10 | - | - | 3.9(0.6) | 5.0 |
| 11 | - | - | 4.4(0.7) | 6.6 |
| 12 | - | - | 5.1(0.6) | 3.1 |
| 13 | - | - | 5.5(1.8) | 3.9 |
| 14 | - | - | 5.9(2.8) | 4.1 |
| 15 | - | - | 6.0(2.2) | 6.6 |

TABLE GII. Parameters for the Lorentz oscillators given by **Eq. (1)** for the ordinary dielectric function $|\tilde{\varepsilon}_{33}|$ for TaAs. $\varepsilon^{\infty} = 1.7$ (Expt.) and 1.9 (DFT), $A_{UV} = 0.1$ (Expt.) and 49.8 (DFT), $E_{UV} = 6.5$ (Expt.) and 6.9 (DFT), $A_{IR} = 6.6$ (Expt.).

| Oscillators (n) | Expt. $E_n$ ($\Gamma_n$) (eV) | Expt. $A_n$ (a.u.) | DFT $E_n$ ($\Gamma_n$) (eV) | DFT $A_n$ (a.u.) |
|---|---|---|---|---|
| 1 | 1.2(0.5) | 8.0 | 0.3(0.4) | 12.4 |
| 2 | 1.8(0.8) | 4.6 | 0.4(0.4) | 14.9 |
| 3 | 2.2(0.2) | 1.6 | 0.8(0.2) | 2.4 |
| 4 | 2.8(0.3) | 0.8 | 1.1(0.6) | 15.6 |
| 5 | 3.0(4.4) | 12.4 | 1.7(0.5) | 8.3 |
| 6 | 4.1(0.8) | 2.8 | 2.0(0.4) | 6.5 |
| 7 | 6.2(4.1) | 5.0 | 2.4(0.9) | 13.5 |
| 8 | - | - | 2.8(0.3) | 2.9 |
| 9 | - | - | 3.0(0.5) | 6.1 |
| 10 | - | - | 3.2(0.3) | 2.3 |
| 11 | - | - | 3.5(0.6) | 5.4 |
| 12 | - | - | 4.0(0.7) | 10.5 |
| 13 | - | - | 4.5(0.5) | 4.3 |
| 14 | - | - | 5.1(0.3) | 0.7 |
| 15 | - | - | 5.6(2.0) | 13.3 |
| 16 | - | - | 5.7(0.6) | 0.9 |

TABLE GIII. Parameters for the Lorentz oscillators given by **Eq. (1)** for the ordinary dielectric function $|\tilde{\varepsilon}_{11}|$ for NbAs. $\varepsilon^\infty = 1.4$ (Expt.) and 1.4(DFT), $A_{UV} = 1.4$ (Expt.) and 9.0 (DFT), $E_{UV} = 6.7$ (Expt.) and 6.4(DFT), $A_{IR} = 12.2$ (Expt.).

| Oscillators (n) | Expt. $E_n$ ($\Gamma_n$) (eV) | Expt. $A_n$ (a.u.) | DFT $E_n$ ($\Gamma_n$) (eV) | DFT $A_n$ (a.u.) |
|---|---|---|---|---|
| 1 | 1.1(0.2) | 11.3 | 0.2(0.3) | 389.3 |
| 2 | 1.7(2.4) | 17.3 | 0.5(0.3) | 29.2 |
| 3 | 3.0(0.9) | 3.4 | 0.8(0.4) | 15.3 |
| 4 | 4.4(1.5) | 2.6 | 1.2(0.4) | 10.4 |
| 5 | 5.9(5.6) | 5.1 | 1.4(0.6) | 7.1 |
| 6 | - | - | 1.8(0.5) | 4.6 |
| 7 | - | - | 2.2(1.5) | 8.6 |
| 8 | - | - | 2.9(0.4) | 4.6 |
| 9 | - | - | 3.1(0.3) | 2.0 |
| 10 | - | - | 3.8(0.5) | 1.7 |
| 11 | - | - | 4.1(0.4) | 4.0 |
| 12 | - | - | 4.6(0.6) | 4.6 |
| 13 | - | - | 5.4(0.3) | 1.6 |
| 14 | - | - | 5.7(1.3) | 4.9 |

TABLE GIV. Parameters for the Lorentz oscillators given by **Eq. (1)** for the ordinary dielectric function $|\tilde{\varepsilon}_{33}|$ for NbAs. $\varepsilon^{\infty} = 2.3$ (Expt.) and 2.0 (DFT), $A_{UV} = 3.4$ (Expt.) and 48.4 (DFT), $E_{UV} = 6.7$ (Expt.) and 6.8 (DFT), $A_{IR} = 8.3$ (Expt.).

| Oscillators (n) | Expt. $E_n$ ($\Gamma_n$) (eV) | Expt. $A_n$ (a.u.) | DFT $E_n$ ($\Gamma_n$) (eV) | DFT $A_n$ (a.u.) |
|---|---|---|---|---|
| 1 | 1.5(1.2) | 20.1 | 0.1(0.2) | 66.6 |
| 2 | 1.9(0.2) | 1.9 | 0.6(0.1) | 3.5 |
| 3 | 2.6(1.5) | 10.2 | 0.8(0.3) | 10.8 |
| 4 | 3.5(0.7) | 3.0 | 1.1(0.5) | 12.8 |
| 5 | 4.6(2.8) | 4.7 | 1.5(0.6) | 11.7 |
| 6 | 5.5(0.8) | 3.5 | 1.8(0.5) | 11.6 |
| 7 | 6.4(1.0) | 2.9 | 2.1(0.5) | 8.9 |
| 8 | - | - | 2.4(0.5) | 10.5 |
| 9 | - | - | 2.8(0.5) | 7.7 |
| 10 | - | - | 3.2(0.5) | 7.9 |
| 11 | - | - | 3.6(0.7) | 6.9 |
| 12 | - | - | 4.1(0.7) | 7.1 |
| 13 | - | - | 4.6(0.7) | 4.9 |
| 14 | - | - | 5.1(0.6) | 5.2 |
| 15 | - | - | 5.4(0.6) | 9.7 |
| 16 | - | - | 6.0(0.7) | 10.5 |